\documentclass[slac_one]{revtex4}
\usepackage[pdftex]{graphicx}
\usepackage{fancyhdr}
\usepackage{atlasphysics}
\begin{document}

\title{{\small{Hadron Collider Physics Symposium (HCP2008),
Galena, Illinois, USA}}\\ 
\vspace{12pt}
Top Quark Physics at the LHC} 

%

\author{Akira Shibata on behalf of the ATLAS and CMS Collaborations}
\affiliation{New York University, 4 Washington Place, NY 10012, USA}

\begin{abstract}
An overview of the prospects of top quark physics at the LHC is presented. The ATLAS and the CMS detectors are about to produce a large amount of data with high top quark contents from the LHC proton-proton collisions. A wide variety of physics analyses is planned in both experiments, and a number of useful insights have already been obtained regarding their detector performance and physics potential. This summary is based on the talk presented at the Hadron Collider Physics Symposium 2008, Galena, Illinois, May 27-31, 2008.
\end{abstract}

\maketitle

\thispagestyle{fancy}

%
%
\section{Introduction}
\subsection{Top Physics timeline at the LHC}

Even though its existence was predicted a few decades ago, and its discovery was made more than a decade ago, the top quark still plays a major role in the forefront of the experimental and theoretical challenges of high energy physics. At the LHC, the importance of the top quark can be recognized in a number of perspectives over the lifetime of the accelerator and the general purpose detectors on the ring, ATLAS and CMS. They are planned to commence operation later this year. In the past years, the two collaborations studied various aspects of the top quark physics using Monte Carlo event generators and detector simulation, not only testing the existing analysis methods developed at the Tevatron experiments, but suggesting a number of new ideas that have shown to be relevant and feasible at the LHC. This paper provides a brief review of these analyses and shows how they fit together in each phase of data-taking operation.  The results were collected mainly from the Computing System Commissioning analysis in the ATLAS experiment \cite{ATLASDET,ATLASCSC} and the Technical Design Report from CMS \cite{CMSDET,CMSPHYSTDR} with updates where available.

\subsection{Top physics: topics of interest and constraints at the LHC}

\begin{figure}[htb]
    \begin{center}
	\includegraphics[width=7cm]{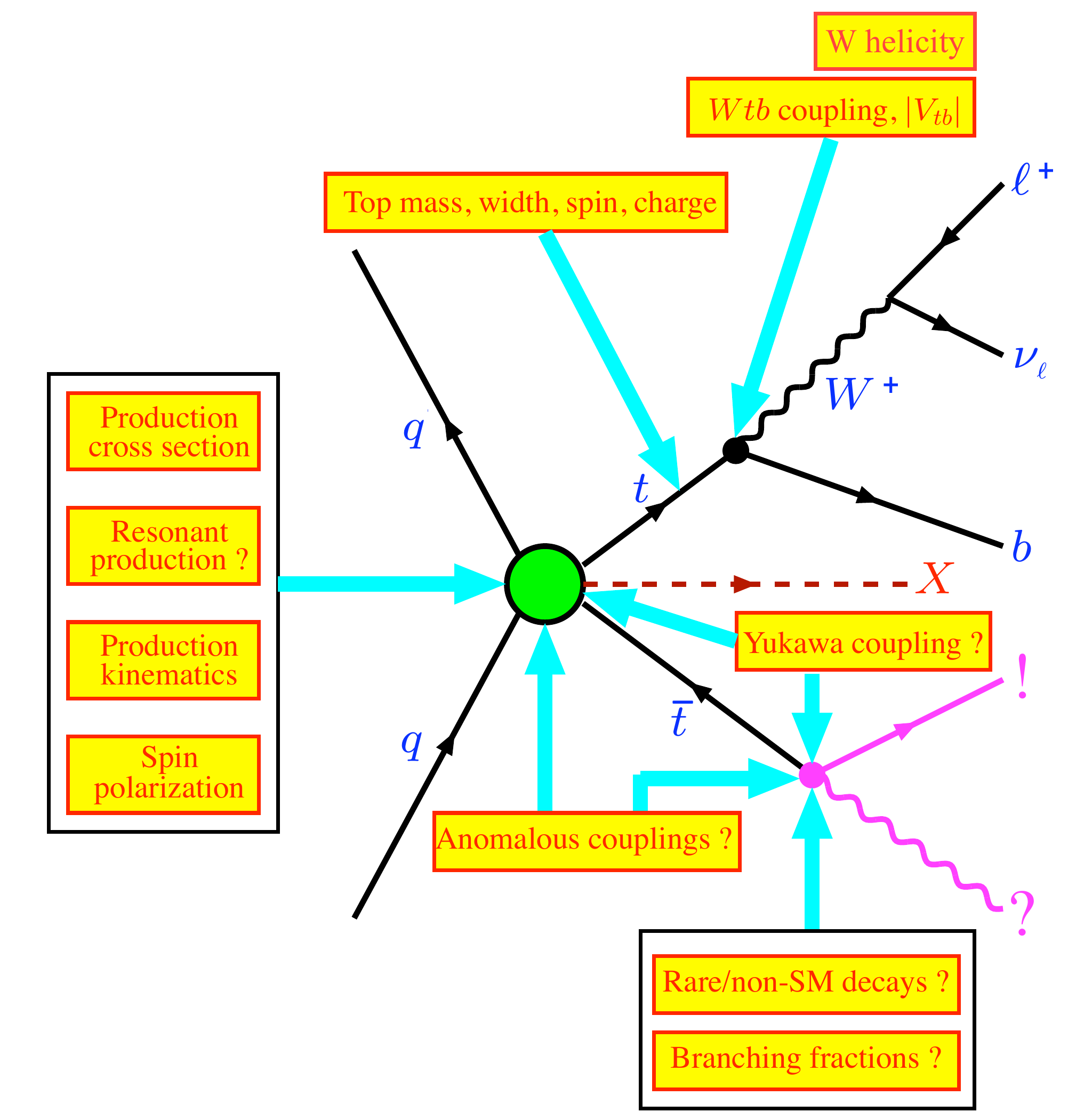} \hspace{1cm}
	\includegraphics[width=7cm]{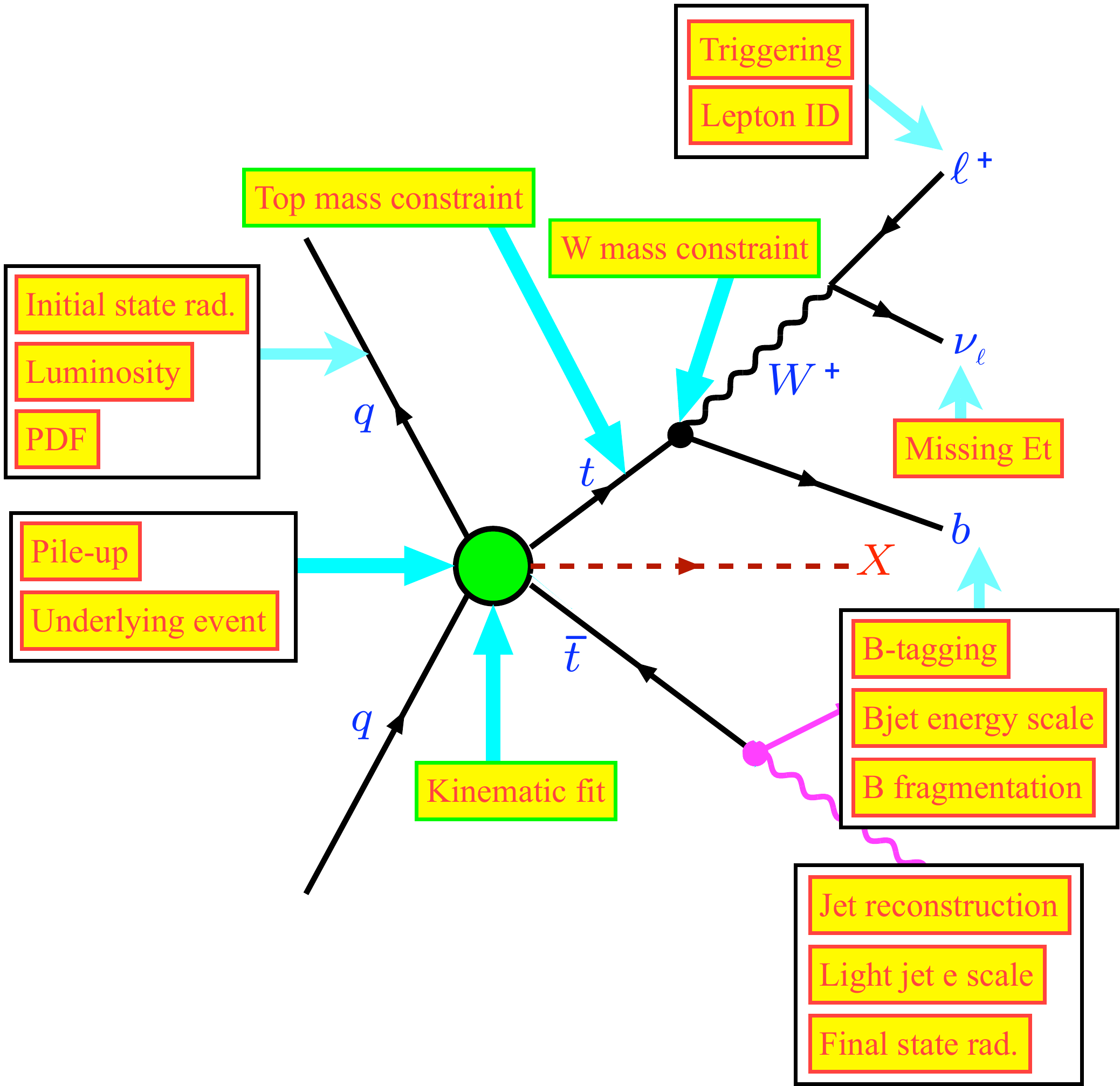} 
	\caption{Diagrams illustrating the theoretical interests (left) and the experimental issues (right) related to top quark physics. The arrows indicate connection between the topics and the objects within \ttbar\ events.}
	\label{Fig::interest}
	\end{center}
\end{figure}

There is a wide variety of accessible top physics topics that are of interest to experimentalists and theorists, using the LHC data. Some of them are illustrated in Figure \ref{Fig::interest}. The production mechanism of the top quark at the highest energy regime of our reach is of the primary interest to us. The measurement of the total production cross-section will enable us to distinguish a number of theoretical predictions by itself. Using additional variables and studying differential cross-sections, we can further study the nature of the production vertex to search for existence of resonant structure or anomalous couplings. The properties of the top quark itself need to be measured as well. This includes mass, width, spin, charge and couplings to other particles including the Yukawa coupling to the Higgs boson. The top quark decays almost exclusively to a $W$ boson and a $b$ quark, though with a good amount of data we can probe the precise nature of the \Vtb\ vertex and possibly an existence of rare decays. On the other hand, extracting such measurements is a challenging endeavor: events produced at hadron colliders are known to have a high multiplicity. Understanding the beam luminosity can be a tricky task. At high luminosity, estimation of pile-up will become a primary concern. The contribution of non perturbative QCD processes to the observed events has large uncertainties too. The parton contents of the beam (PDF), the amount of initial state radiation (ISR) and of the accompanying underlying event have to be taken into account and understood before the experiments can carry out precision measurements.

Several experimental uncertainties affect the reconstruction of the events: a sound measurement of the jet energies can only be achieved after extensive calibration efforts.
The missing transverse energy (\met) measurement requires a global understanding of the detector. Care needs to be taken to avoid possible bias from trigger requirements to the measured quantities. In addition, the understanding of the $b$-tagging performance is a complex task in itself and can be affected by theoretical issues such as the $b$-fragmentation too.

During the lifetime of the experiment the analysis methods will evolve: early data analysis will use the most reliable information only. At this point known physics processes can be used to calibrate and better understand the detector in order to be able to perform more complex analyses. This applies to the case of top quark physics:
its observation will be challenging since it typically requires jets and \met, however once observed, the \ttbar\ event topology has a number of applications that will help us understanding the detector.

\subsection{Tevatron and the LHC}
The LHC is literally a top factory. Its large center-of-mass energy, $\sqrt{s}=14$ TeV, will yield the \ttbar\ production cross-section of 833 pb, larger than at the Tevatron, $\sqrt{s}=1.96$ TeV, by a factor of $\sim$100. The design luminosity of the LHC is also a factor of 100 larger than the Tevatron luminosity, so statistics will quickly become a non-issue for many top quark measurements. At its design luminosity, one \ttbar\ pair is produced almost every second at the LHC as opposed to 10 pairs per day at the Tevatron. 

In addition to the production rate, the \ttbar\ production mechanism differs significantly. The proton-antiproton collisions at the Tevatron produce \ttbar\ pairs predominantly (85\% of the cases) through quark-antiquark annihilation, while the majority of the \ttbar\ production at the LHC (90\% of the cases) is due to gluon-gluon fusion. At the LHC, the PDF uncertainty for gluon-gluon fusion process is much smaller than at the Tevatron while $q\bar{q}$ processes entail larger uncertainty. The total uncertainty on the \ttbar\ cross-section due to PDF is estimated to be smaller at the LHC \cite{Huston2007, Cacciari2008}. 

\begin{figure}
	\begin{center}
	\includegraphics[width=12cm]{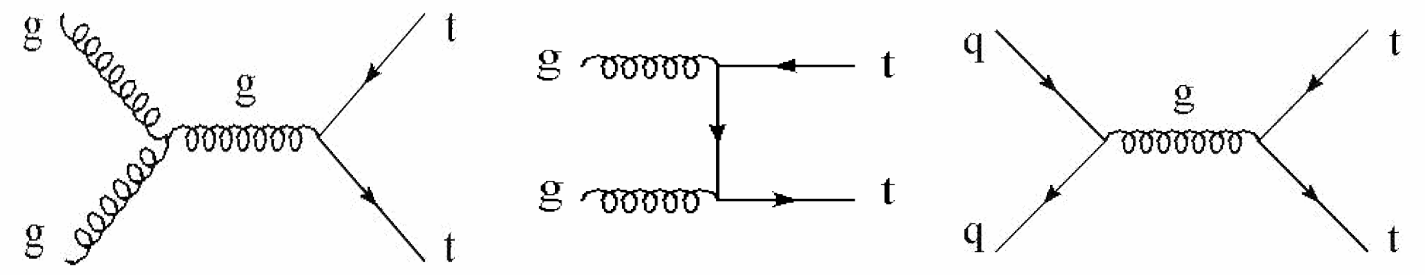} 
	\caption{Some of the main \ttbar\ production diagrams. Gluon-gluon fusion diagrams (left two) are the main production mode at the LHC while the quark-antiquark annihilation (right) is dominant at the Tevatron.}
	\end{center}
\end{figure}

\subsection{ATLAS and CMS}
The ATLAS and the CMS detectors were built primarily to discover the Higgs boson. Luckily, the energy scale for their performance optimization is highly desirable for top quark observation as well. The design of the two detectors is similar in that they have a cylindrical structure with inner tracking detectors, calorimeters, and muon systems with superconducting magnets in between. Many sub-detectors are currently being commissioned and calibrated with cosmic ray muons.

In addition to detector commissioning, the commissioning of the computing facility is an ongoing effort in both collaborations. 

\subsection{The LHC expectations in the first year}
According to the current schedule, the LHC accelerator will start operation in 2009 with center-of-mass energy less than the design beam energy (such as 5 TeV + 5 TeV) to avoid time-consuming commissioning required to achieve energy above 5.5 TeV. The initial luminosity will be lower than the design luminosity by a factor of 1000. The lower beam energy means that the \ttbar\ cross-section will be reduced by half. This applies roughly to all background processes as well. Even with 3 months of uninterrupted data-taking, the integrated luminosity will be less than 100 pb$^{-1}$. The number of \ttbar\ events will therefore be less than 40,000 in the first year of operation. The changes in the initial beam parameters were determined only shortly before this review was compiled. All studies presented here assume the design beam energy.

\section{Early data analysis - Physics commissioning and first measurements}
As explained earlier, at the beginning of the data taking the detector performance
will need to be understood and therefore the top quark observation will not be the
first measurement to be done at the LHC. However, the first observation of the top quark will be an important milestone indicating that physics analysis environment is well established. Thus, a number of analyses are in preparation aiming for the top observation with the initial data, followed by cross-section measurements. 

\subsection{Dilepton channel}

\begin{figure}[htb]
	\begin{center}
	\includegraphics[width=8cm]{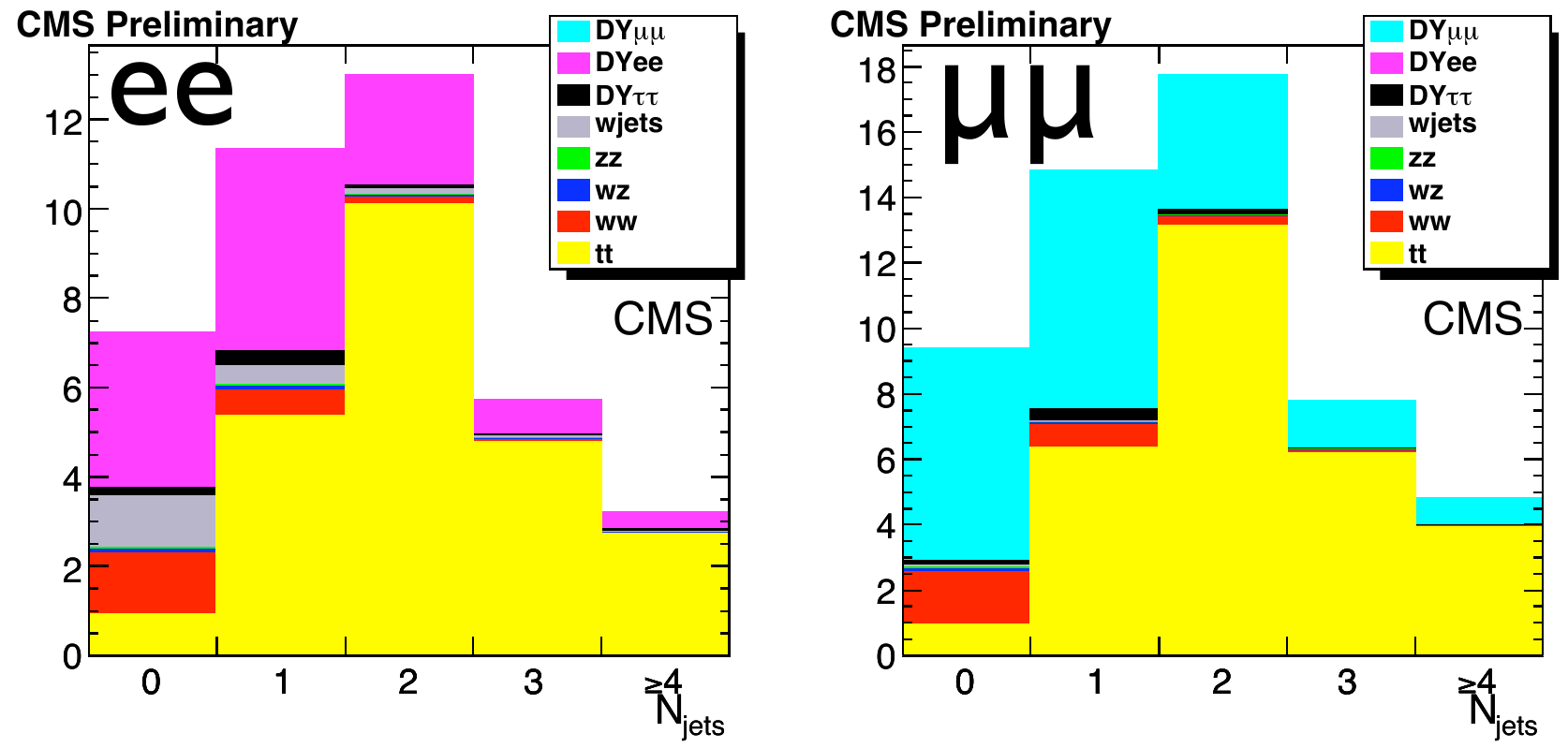} 
	\includegraphics[width=8cm]{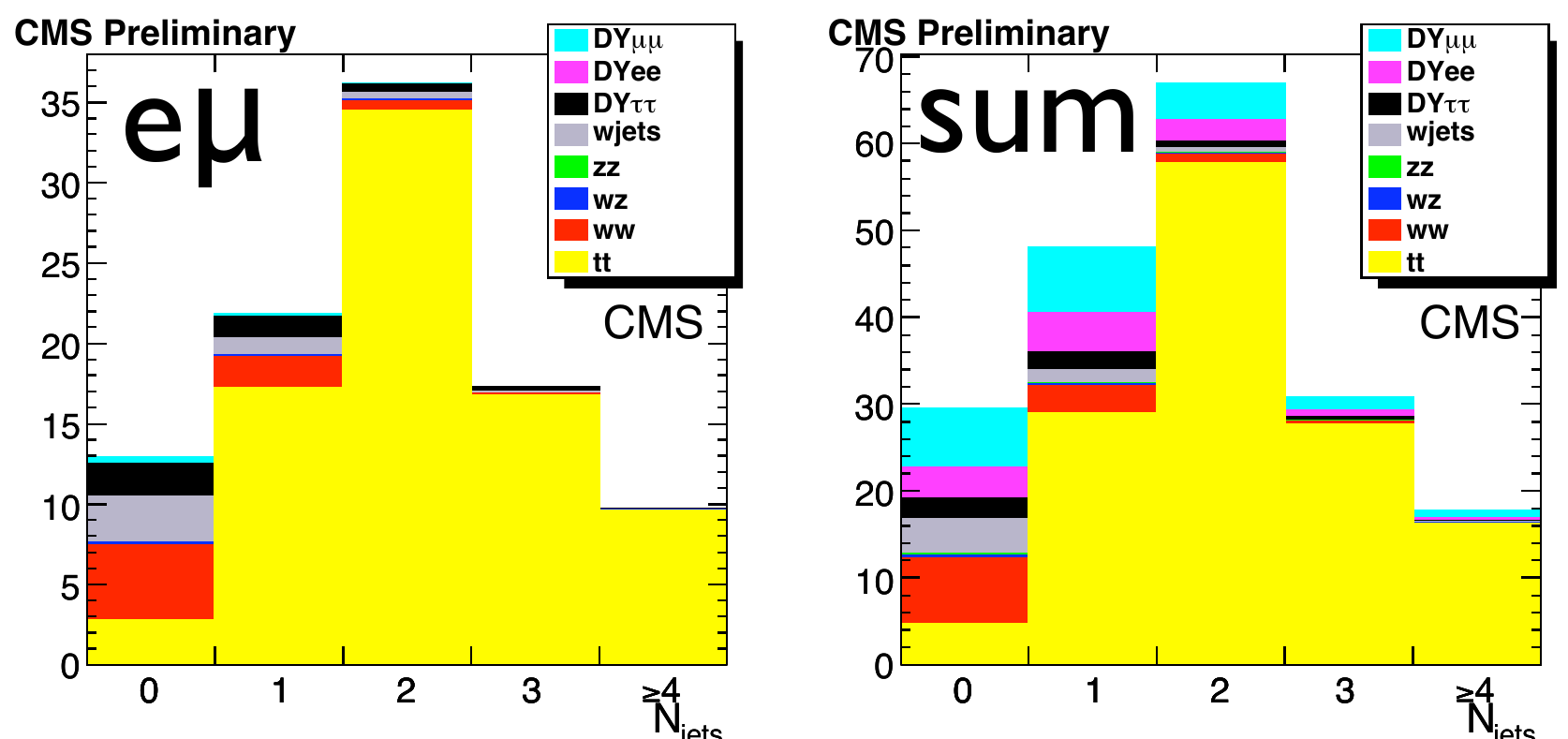} 
	\caption{Jet multiplicity for jets with \pt\ $>$ 30 GeV, normalized to 10 pb$^{-1}$.}
	\label{Fig::Dilepton}
	\end{center}
\end{figure}

The dileptonic final state is a rare signature where both $W$ bosons, produced by the decay of the top pair, decay leptonically (electron or muon), and can be triggered with very high efficiency. Although the cross-section for dileptonic channels is small, this is possibly the first place where the evidence of top events can be seen. The lepton fake contribution from QCD jets will be small thanks to the requirement the presence of two leptons in the final state. Therefore the main background comes from Drell-Yan (DY) production and diboson processes. In events with an electron and a muon in the final state, even DY events will not contribute to the background. Figure \ref{Fig::Dilepton} shows the jet multiplicity in dilepton final states assuming 10 pb$^{-1}$ of data, as obtained in a CMS analysis \cite{CMSPASTOP-08-001}. A \pt\ threshold of 16 GeV and 17 GeV was used to trigger single electrons and muons respectively. The offline lepton selection required two opposite charge leptons with isolation criteria based on the calorimeter and the tracking. Further event selection simply removed events with low missing \et\ events and $Z$ events were removed by applying a window cut on the dilepton mass. The \ttbar\ signal can be seen clearly in the figure. The signal purity in $e\mu$ events is outstanding. It was estimated that, by counting the number of events with two or more jets, the cross-section can be measured with a $~13\%$ statistical uncertainty and a systematic uncertainty of the same order using 10 pb$^{-1}$ of data.

\subsection{Semileptonic channel}
\begin{figure}[htb]
	\begin{center}
	\includegraphics[width=7cm]{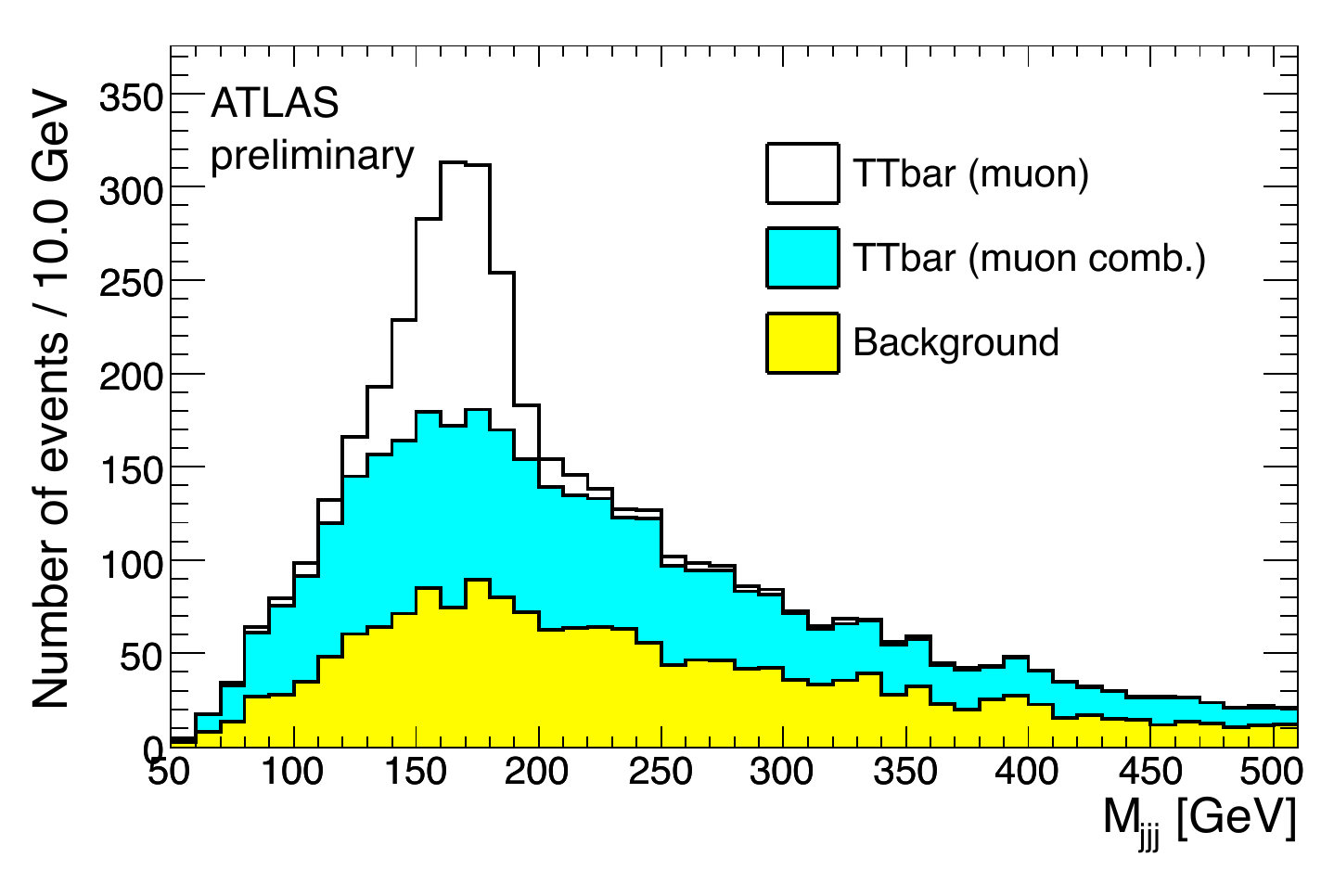}  \hspace{1cm}
	\includegraphics[width=7cm]{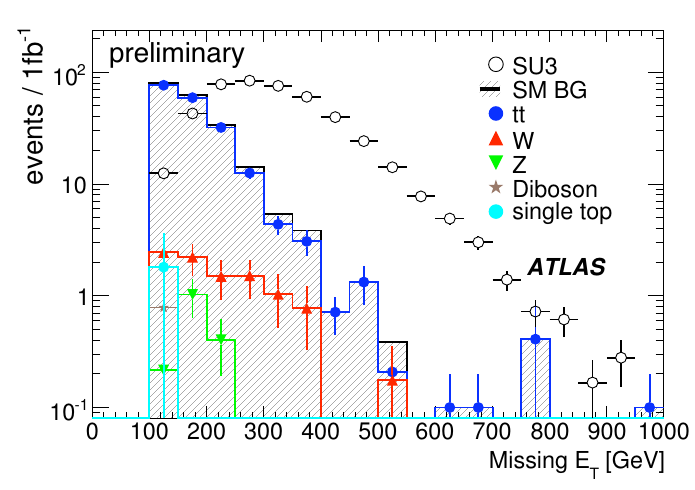} 
	\caption{Left: mass of the highest \pt\ trijet combination before applying $W$ mass window cut. Right: \met\ spectrum compared with a SUSY signal and other potential background to SUSY.}
	\label{Fig::Semilep}
	\end{center}
\end{figure}

Measurements in the semileptonic (or ``lepton plus jets'') channels benefit from large cross-section (30\% of the \ttbar\ events decay into $e/\mu$+jets) and in this channel the top quark can be fully reconstructed from its hadronic decay products. Combinatorial ambiguity can be large as shown in Figure \ref{Fig::Semilep} (left), though a clear top mass peak is visible. B-tagging can reduce this combinatorial background significantly but is not used in the early-data analysis by ATLAS \cite{ATLASCSC}. Instead, top quarks are reconstructed by taking the highest \pt\ trijet combination assuming that the jets from the top decay tend to be collimated in a similar direction. To increase purity, it is required that there is at least one dijet combination with invariant mass close to the mass of the $W$ boson. With the requirement of \met$>$20 GeV and an isolated electron or muon with \pt$>$25 GeV, the selection efficiency is 18.2\% and 23.6\% respectively. This is halved by requirement of $W$ mass. The background can be estimated by fitting a Chebyshev polynomial function while obtaining the signal yield using a Gaussian. Precision of $\Delta \sigma/\sigma=7~(stat.)~\pm~15~(syst.)~\pm~3~(PDF)~\pm~5\%~(lumi)$ was estimated at 100 pb$^{-1}$. Although this data-driven method is robust against theoretical uncertainties, it is sensitive to the stability of the fit. An alternative counting method that uses Monte Carlo $W$ + jets sample is suggested. This method is sensitive to uncertainty on the rate of additional jets, but a competitive precision of $\Delta \sigma/\sigma=3~(stat.)~\pm~16~(syst.)~\pm~3~(PDF)~\pm~5\%~(lumi)$ has been obtained, and it can serve as a cross check. An improvement to this analysis is to use a data-driven method to estimate the $W$ + jets background, which is currently under study.

\section{Top as background to new physics}
The measurement of the \ttbar\ cross-section is important since the NLO theoretical cross-section is highly sensitive to the QCD scale variation, estimated to be $\sim$12\% \cite{Kidonakis2003}. There are ongoing efforts to complete full NNLO calculation in a near future, presumably reducing the uncertainty by a half. While LHC experiments are challenged to test the theory at this precision, the measurement is an important input to new physics searches, many of which seek for a final state topology much similar to top events. This includes among the others some Higgs searches (e.g. $H\to\tau\tau, WH, t\bar{t}H$), supersymmetry searches and the twin Higgs model (e.g. $W_H\to tb$) among others.

Many supersymmetry searches look for a large \met\ accompanied by lepton(s) and jets. Signal events typically have much more missing energy than \ttbar\ though the estimation of the background coming from the tail of the distribution is crucial to identify the signal with confidence \cite{ATLASCSC}. This is illustrated in Figure \ref{Fig::Semilep} (right). The \met\ tail cannot be estimated reliably from detector simulation, and a number of data-driven methods are being tested. Such method requires a well-understood control region rich in background but sparse in signal. For this reason, one proposed method relies on well reconstructed top events to estimate the contamination in the signal region \cite{ATLASCSC}.

Obtaining a suitable control region can be problematic depending on the signal model sought after. Signatures of some SUSY parameter space can overlap largely with \ttbar\ such as SU4. In this scenario, the event yield from SUSY in the above-mentioned semileptonic \ttbar\ analysis can be as large as one fifth of \ttbar. It is therefore crucial to measure \ttbar\ cross-section by a variety of methods to ensure that consistency is observed in a largest possible phase space.

\section{Top as a candle in the dark}
Once the first observation of the top quark is established with first few fb$^{-1}$ of data, and the consistency with the Standard Model is confirmed, \ttbar\ events can serve as a standard reference point. New methods have been developed that are useful to deal with detector performance issues that are otherwise difficult to understand. 

\subsection{Efficiency of $b$-tagging}
Since a good number of \ttbar\ events can be obtained with a reasonable purity without using $b$-tagging, such a sample can be used to measure the $b$-tagging performance. One conventional method uses dijet events in which one of the jets is tagged using soft lepton tag. 
However the $b$-tagging performance might vary in events with higher \pt\ jets and larger jet multiplicity. To extract the $b$-tagging efficiency in such environment \ttbar\ events can be used, even if the production rate is much smaller than dijet events.

One simple method is to count the number of events with different $b$-tagged jet multiplicity \cite{ATLASCSC}. To obtain the efficiency, the multiplicity distribution estimated from Monte Carlo simulation including expected background events is fitted to the observed distribution. This method relies on detector simulation and only the integrated efficiency can be measured. Precision of $\sim$5\% is expected with 100 pb$^{-1}$ of data. 

\begin{figure}
	\begin{center}
	\includegraphics[height=5.5cm]{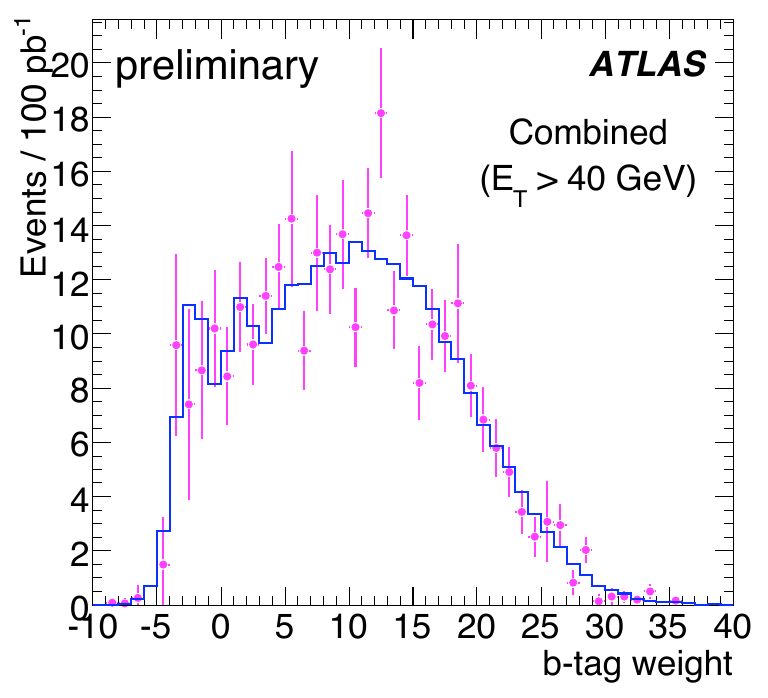} \hspace{1cm}
	\includegraphics[height=5.75cm]{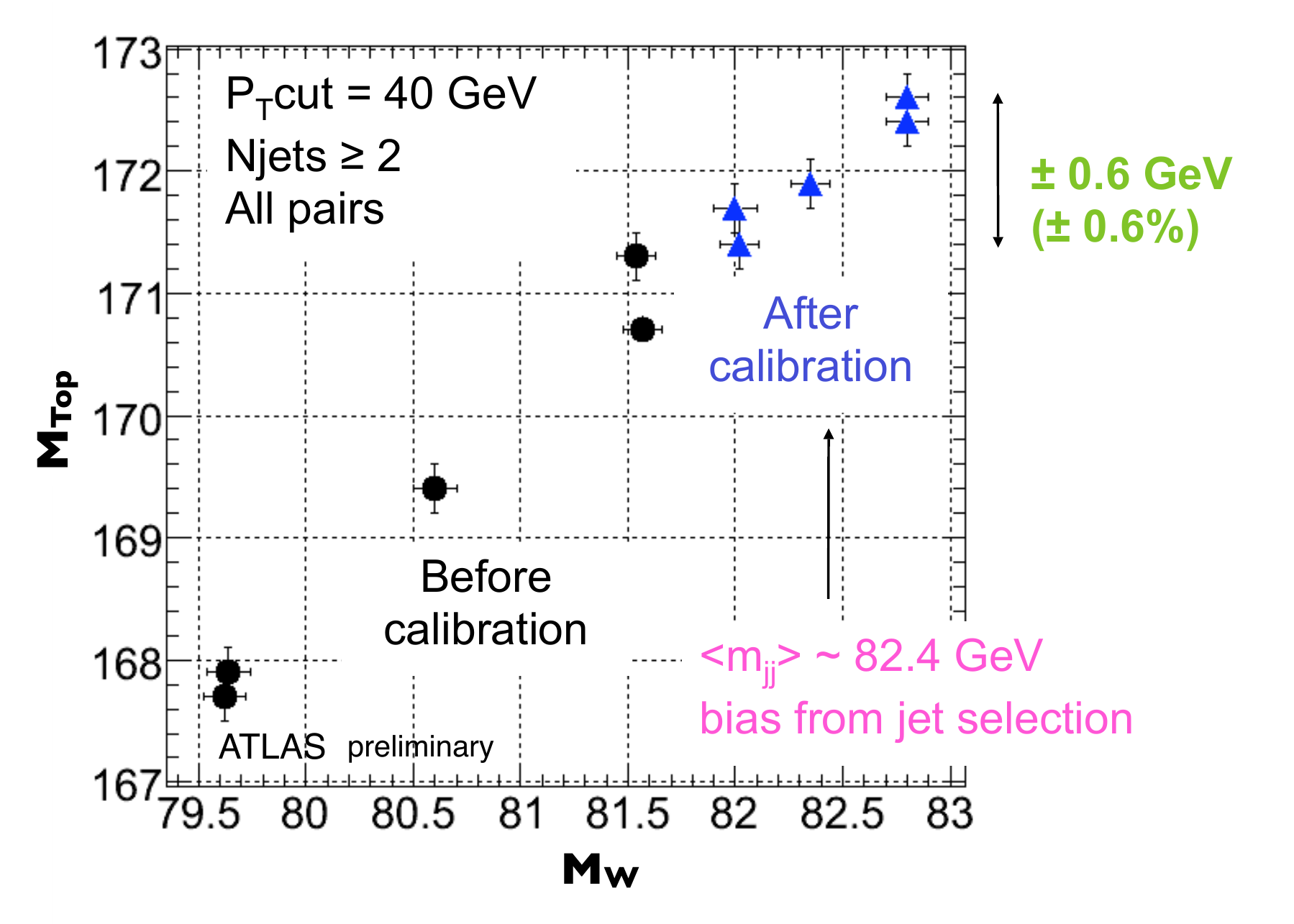} 
	\caption{Left: The $b$-tagging discriminant variable distribution estimated using enriched $b$-jet sample from \ttbar\ events at 100 pb$^{-1}$ (points) and the true distribution (line). Right: reconstructed top and $W$ mass as a result of ensemble testing at 1 fb$^{-1}$ before and after the in-situ calibration.}
	\label{Fig::btag}
	\end{center}
\end{figure}

Another method tries to identify a pure sample of $b$-jets by exploiting the \ttbar\ event topology such as the reconstructed top mass. However, due to the existence of combinatorial and other physics backgrounds, one can only extract $b$-tagging performance on a statistical basis subtracting the background distribution from the histogram. Once background is properly subtracted, it is then possible to obtain the shape of $b$-tagging discriminant variable. With it, $b$-tagging efficiency can be measured as a function of the selection criteria as shown in Figure \ref{Fig::btag} (left). This method is statistically less robust compared to the first method, though with enough data, it can be extended to include other variables such as \pt\ and $\eta$.

\subsection{Jet energy scale}
The jet energy calibration is crucial for almost all analyses at the LHC including the top mass measurement. However, the precise measurement of jet energy scale (JES) and jet energy resolution is known to be a non-trivial task. In \ttbar\ events, one can use the non-$b$-tagged jets from the top decay since they are known to originate from the light quarks from the $W$ decay. The current method uses a number of template models with varying scale and resolution \cite{ATLASCSC} as well as a kinematic fit on the \ttbar\ event topology to purify the candidate jets \cite{CMSTOPJES07}. The jet energy scale can be obtained by fitting the templates to the data. This calibration can be performed in-situ in semileptonic \ttbar\ events and can improve the stability of top mass measurement significantly as shown in Figure \ref{Fig::btag} (right).

There is a remaining issue with the jet energy scale when the jet originates from a $b$ quark. In this case an additional correction is necessary to take into account a
larger fraction of energy loss outside the jet cone than in the case of light jets.
 A method has been explored that extracts $b$-jet energy scale (BJES) independent of the top mass \cite{Fiedler2007} though it has not been proven to be viable. The current best method is to fix the top mass to the known best value but this is clearly unfavorable.


\section{Top property measurement}
With increasing amount of data accumulated at the LHC after the early data-taking phase, precision measurements of the top quark properties are expected to provide new insights into the electroweak symmetry breaking mechanism. For its heavy mass, the top quark properties may be the first indication of the existence of new physics.

\subsection{Top mass measurement}
\begin{figure}
	\begin{center}
	\includegraphics[width=8cm]{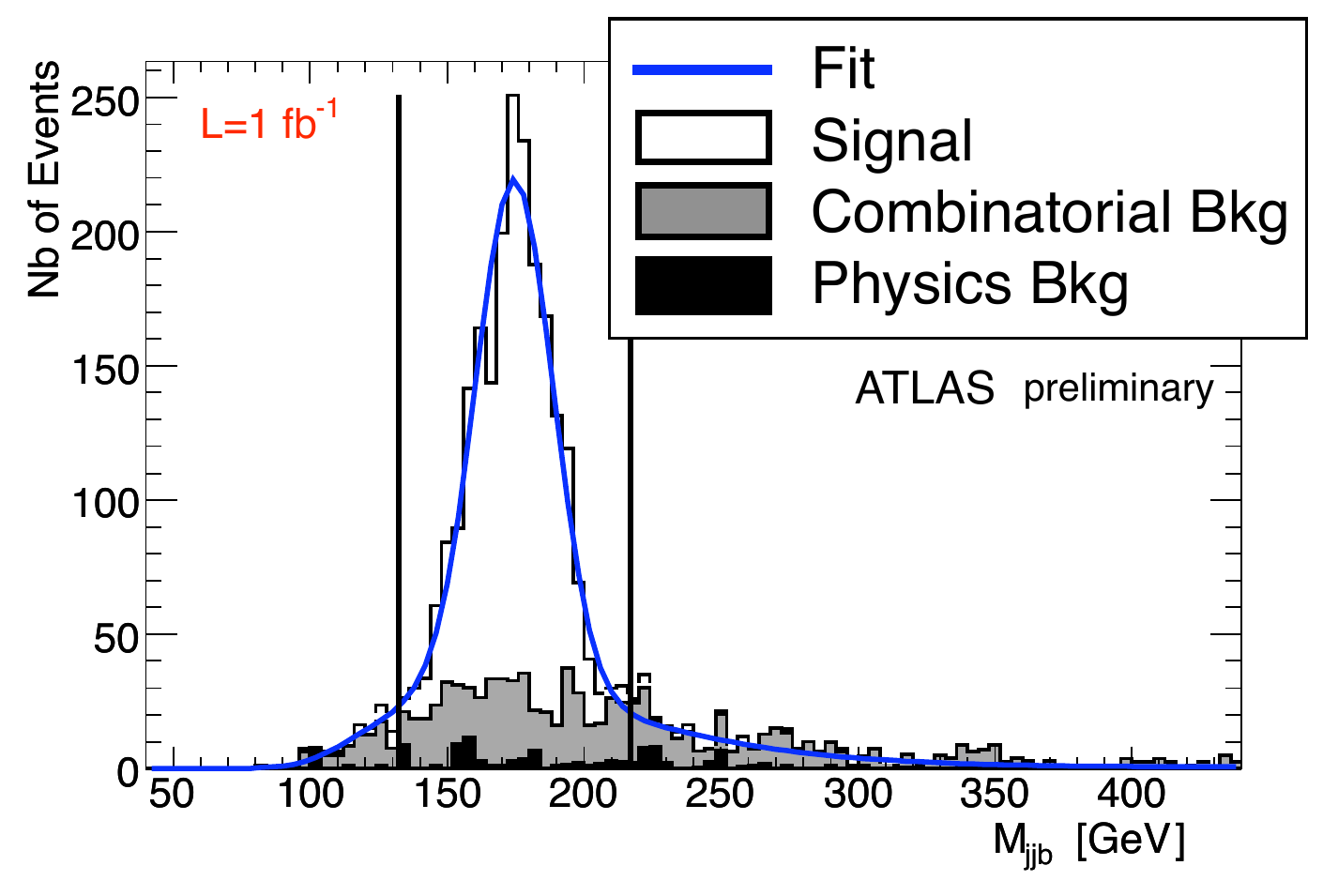} \hspace{1cm}
	\includegraphics[width=5.5cm]{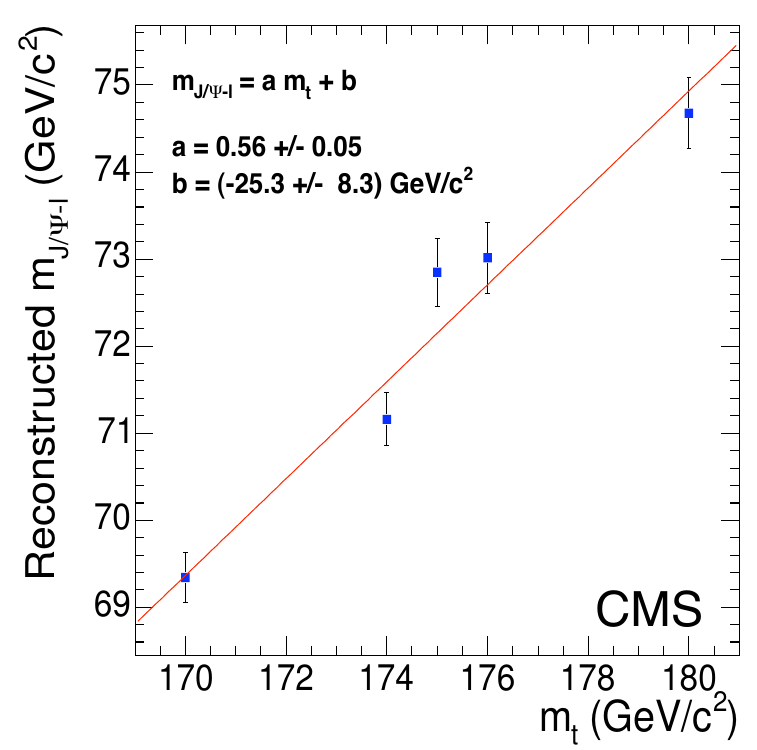} 
	\caption{Left: Reconstructed hadronic top mass distribution in the semileptonic decay mode. Right: the correlation between the tri-lepton invariant mass and the top mass in J/$\psi$ + lepton mode.}
	\label{Fig::mass}
	\end{center}
\end{figure}

One of the most important variable to be measured is the top mass. The precision achieved at the Tevatron is already nearing an impressive 1 GeV \cite{TevEWWG2008}. The LHC experiments will be much less limited in terms of statistics, though the same types of systematic uncertainties will exist, being typical for hadron colliders. A number of mass measurement methods have been studied, some are already used at the Tevatron and some are new. The most prominent ``Golden'' channel is the semileptonic decay mode where the top quark can be clearly reconstructed. Combinatorial and physics background (mainly $W$ + jets) can be reduced with $b$-tagging and tight cuts on the jet \pt. Instrumental background contamination is reduced by requiring large \met\ and an isolated high-\pt\ lepton. A kinematic fit can fully exploit the over constraints that exist in semileptonic \ttbar\ decays, though a simpler method has shown to be as competitive and robust. 
In this method, the top candidate is reconstructed by combining the three jets nearest to each other in angle and purified using the $W$ mass constraint. 
Requiring the invariant mass of the hadronic $W$ and the leptonic $b$-jet to be greater than 200 GeV and one of the lepton and the leptonic $b$-jet to be smaller than 160 GeV, a purity of the sample of 78\% can be reached within 3 sigma from the reconstructed top mass.
The mass distribution after the final selection is shown in Figure \ref{Fig::mass} (left). The corresponding efficiency is 0.82\%. The mass extracted from a Gaussian fit is $174.6~\pm~0.5~(stat.)~\pm~0.2~(per~1\%~JES)~\pm~0.7~(per~1\%~BJES)~\pm~0.4~(ISR/FSR)$ for the input top mass of 175 GeV. The in-situ jet energy calibration mentioned in the previous section yields 1\% precision with 1 fb$^{-1}$ of data but the uncertainty on the $b$-jet energy scale may be much larger.

The mass measurement based on jet energy is sensitive to potentially large energy scale uncertainties as discussed above. A method that relies entirely on the lepton information is suggested by CMS \cite{CMSPHYSTDR}. The method requires semileptonic events where the $b$ quark from the top decay subsequently decays into J/$\psi$ after forming a $B$ meson. The branching ratio for this decay is of the order of $5.5\times10^{-4}$ and it requires a very large amount of data. The extraction of the mass relies on Monte Carlo simulation from which the correlation between the peak of the invariant mass of the three leptons and the mass of the top quark is extracted as shown in Figure \ref{Fig::mass} (right). The statistical uncertainty with 20 fb$^{-1}$ is $\sim$1 GeV and the systematic uncertainty is dominated by the uncertainty of the MC generator parameters, estimated to be 1.5 GeV, using TopRex\cite{Slabospitsky} and Pythia\cite{PYTHIA} generators.

\subsection{\ttbar\ spin correlation}
One interesting property of \ttbar\ production is the spin correlation between $t$ and $\bar{t}$. The double differential cross-section reveals the correlation as follows:
\begin{equation}
\frac{1}{N}\frac{d^2N}{d\cos\theta_ld\cos\theta_q}=\frac{1}{4}(1-\mathcal{A}\kappa_l\kappa_q)\cos\theta_l\cos\theta_q
\end{equation}
and it is observable using the decay products (lepton, l, for leptonic top, light jet, q, for hadronic top) as spin analyzers (with analyzing power $\kappa$) of either top in their helicity basis. Standard Model predicts a correlation $\mathcal{A}$ of 0.32. A good understanding of the bias in measured angles is necessary to interpret the observation and the current method relying on detector simulation estimated precision of 17\% with 10 fb$^{-1}$. Spin correlation becomes particularly relevant if a \ttbar\ resonance state is discovered. Theoretical predictions can be distinguished based on this information, which reflects the spin of the resonant particle.

\section{Single top measurement}
\begin{figure}[tb]
	\begin{center}
	\includegraphics[width=11cm]{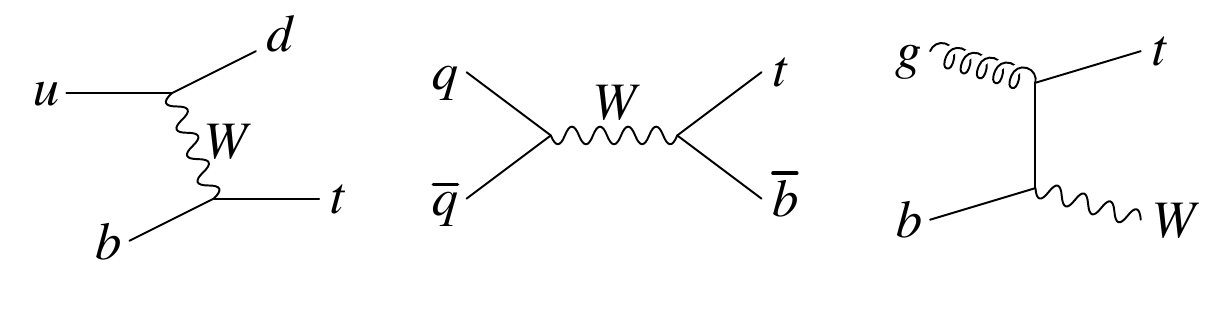} 
	\caption{Feynman diagrams of the leading order single top processes. From left: t-channel, s-channel and $tW$ associated production.}
	\label{Fig::st-feyn}
	\end{center}
\end{figure}

At the LHC the production process of a single top in the final state has a cross-section of about 1/3 of the \ttbar\ production.
Like \ttbar, the production rate of the single top processes will also be larger by a factor of 100 or so compared to the Tevatron. The dominant t-channel process has a cross-section of 244.6 pb, the second largest $tW$-channel of 62.1 pb and the s-channel of 10.65 pb. Figure \ref{Fig::st-feyn} shows the Feynman diagrams of each process. 

The single top production provides vital information about the top quark, which complements our knowledge gained from the \ttbar\ process. It is initiated via the weak interaction between quarks, unlike \ttbar, which is a manifestation of strong interaction. This implies that the study of the single top processes will enable us to test the Standard Model from a different perspective leading to a universal understanding of the particle.

\subsection{t-channel single top}
\begin{figure}[htb]
	\begin{center}
	\includegraphics[height=5cm]{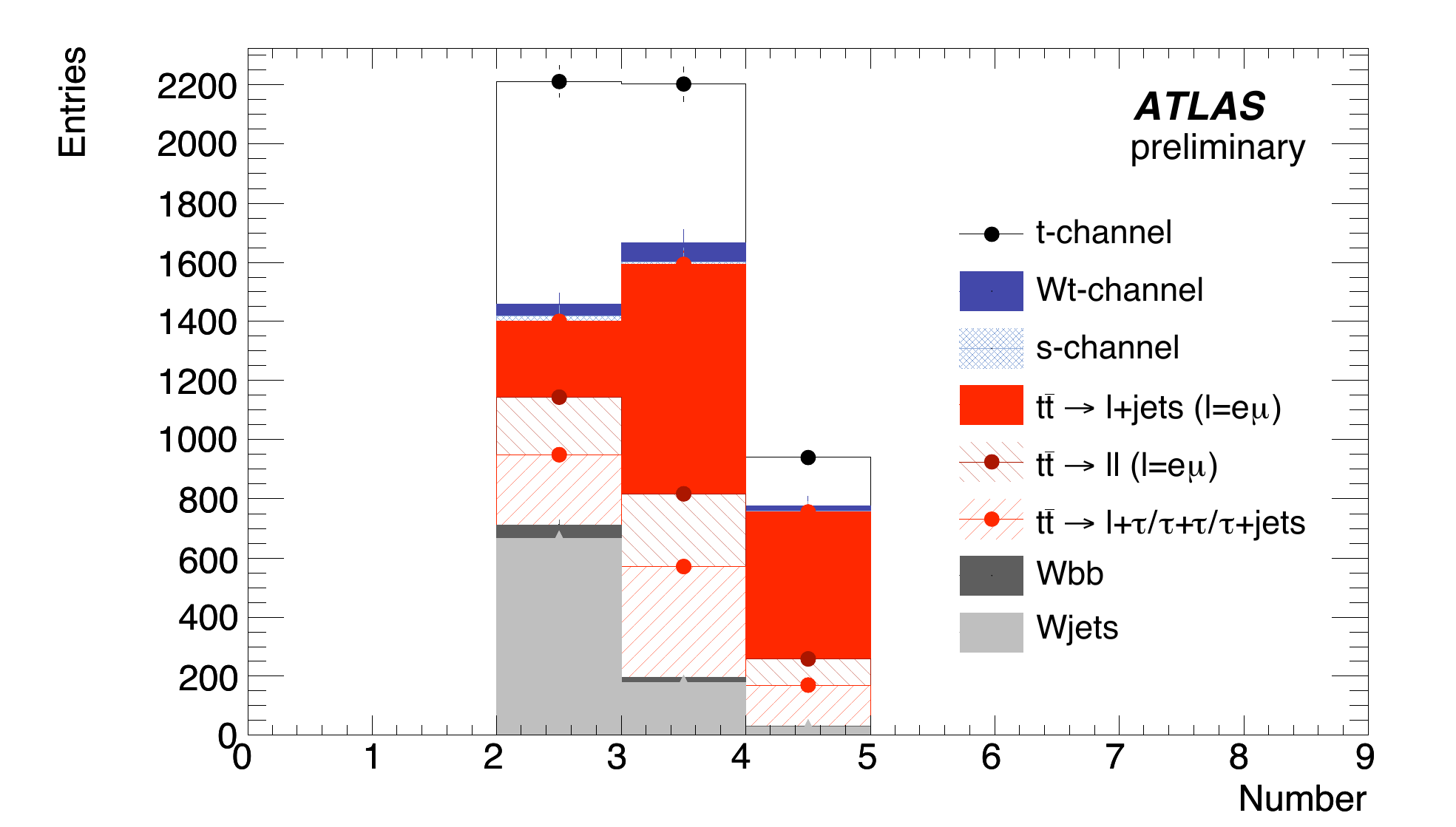} 
	\includegraphics[height=5cm]{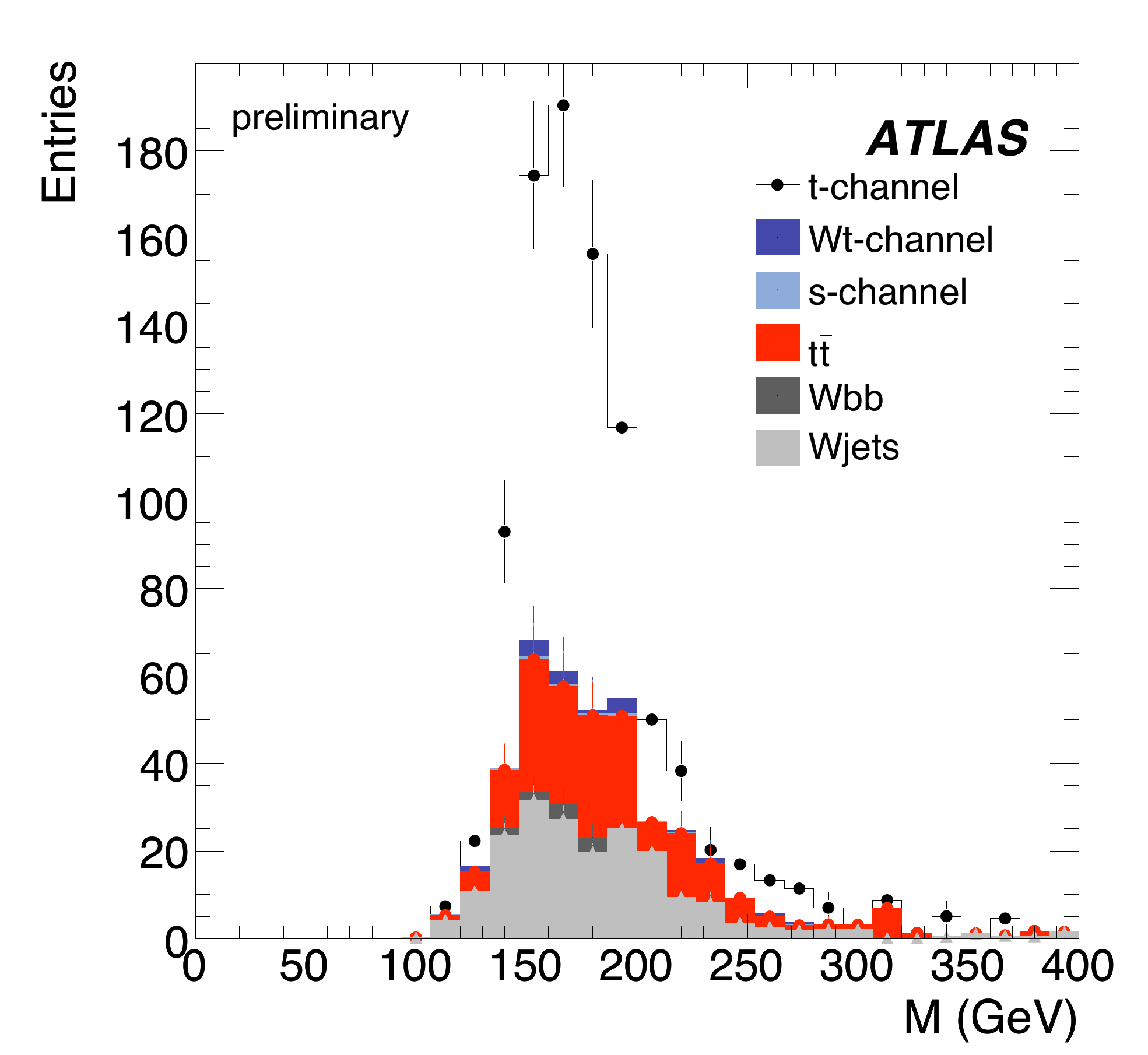}
	\caption{Left: Jet multiplicity distribution of selected t-channel sample. Right: top mass distribution after event selection using Boosted Decision Trees. Both at 1 fb$^{-1}$}
	\label{Fig::st}
	\end{center}
\end{figure}

The t-channel process has by far the largest cross-section and its production rate is roughly proportional to \Vtb, a quantity not directly measured elsewhere. The parity violating V+A coupling of the weak interaction can be tested in this channel by measuring the polarization of the top quark using its decay products. Extracting the signal in the t-channel single top (and in fact in all others) is challenging due to the large background contribution coming from \ttbar\ and $W$ + jets. This is illustrated in Figure \ref{Fig::st} (left). It was shown in the Tevatron analyses that multivariate background rejection techniques are highly effective to purify the single top sample and this has also been demonstrated in ATLAS \cite{ATLASCSC}. In addition to the cut based selection using isolated lepton, \met, $b$-tagging and jet multiplicity cuts, additional 12 variables were combined using the Boosted Decision Tree method. As a result, a clear signal can be seen even with 1 fb$^{-1}$ of data as shown in Figure \ref{Fig::st} (right). With 10 fb$^{-1}$, a 10\% precision on the measured cross-section has been estimated including systematic uncertainties, which translates into a 5\% uncertainty on \Vtb\ measurement.

\subsection{Top-$W$ associated production}
Most current single top studies at the LHC rely heavily on Monte Carlo simulation. Further effort is clearly necessary to reduce this aspect using a more data-driven method like the one found in a CMS analysis \cite{CMSPHYSTDR} shown here. The associated top-$W$ production has a final state extremely similar to \ttbar. To estimate the contamination from this background, a \ttbar\ enriched control region is defined by requiring an additional jet. The ratio of the number of events in the signal and the control regions is calculated from MC for the signal and the \ttbar\ background separately. It is then possible to extract the number of signal events by solving equations relating the number of observed events in each region in terms of these ratios as follows:
\begin{eqnarray}
S & = & \frac{R_{t\bar{t}}(N_s-N_s^0)-(N_c-N_c^0)}{R_{t\bar{t}}-R_{tW}} \\
B & = & \frac{(N_c-N_c^0)-R_{tW}(N_s-N_s^0)}{R_{t\bar{t}}-R_{tW}}+N_s^0
\end{eqnarray}
Where $R$ are the ratios as mentioned above, $N_s$ ($N_c$) is the number of observed events in the signal (control) region and $N_s^0$ ($N_c^0$) is non-\ttbar\ background in those regions. Non-\ttbar\ background is still estimated entirely from MC. 
With 10 fb$^{-1}$ of data, the estimated uncertainties on the measured cross-section of the semileptonic decay mode are $\Delta \sigma/\sigma=\pm7.5\%~(stat.)~\pm16.8\%~(syst.)~\pm15.2\%~(MC~stat.)$.

\section{Top as signature to new physics - \ttbar\ resonance and boosted top}
As well as being background to new physics, the top quarks can themselves be a signal of new physics. Alternative (non-Higgs) models of electroweak symmetry breaking tend to involve resonances that couple strongly to the top quark, and therefore top is often called ``the best probe for EWSB'' in this respect. For example interactions predicted by theories such as \cite{Frederix2007, Fitzpatrick2007, Technicolor}:
\begin{eqnarray}
	pp & \to  X                   \to  t~\bar{t}                   & ~ (Extra~dimensions~with~resonance~X) \\
	pp & \to  b'\bar{b}'          \to  W^-tW^+\bar{t}             & ~ (Extra~generation~b')\\
	pp & \to  \tilde{g}~\tilde{g}  \to  \widetilde{g}t~\widetilde{g}\bar{t} & ~ (Top~color~with~new~gauge~boson~\tilde{g})
\end{eqnarray}
can produce top quarks in the final state.

\begin{figure}
	\begin{center}
	\includegraphics[width=6cm]{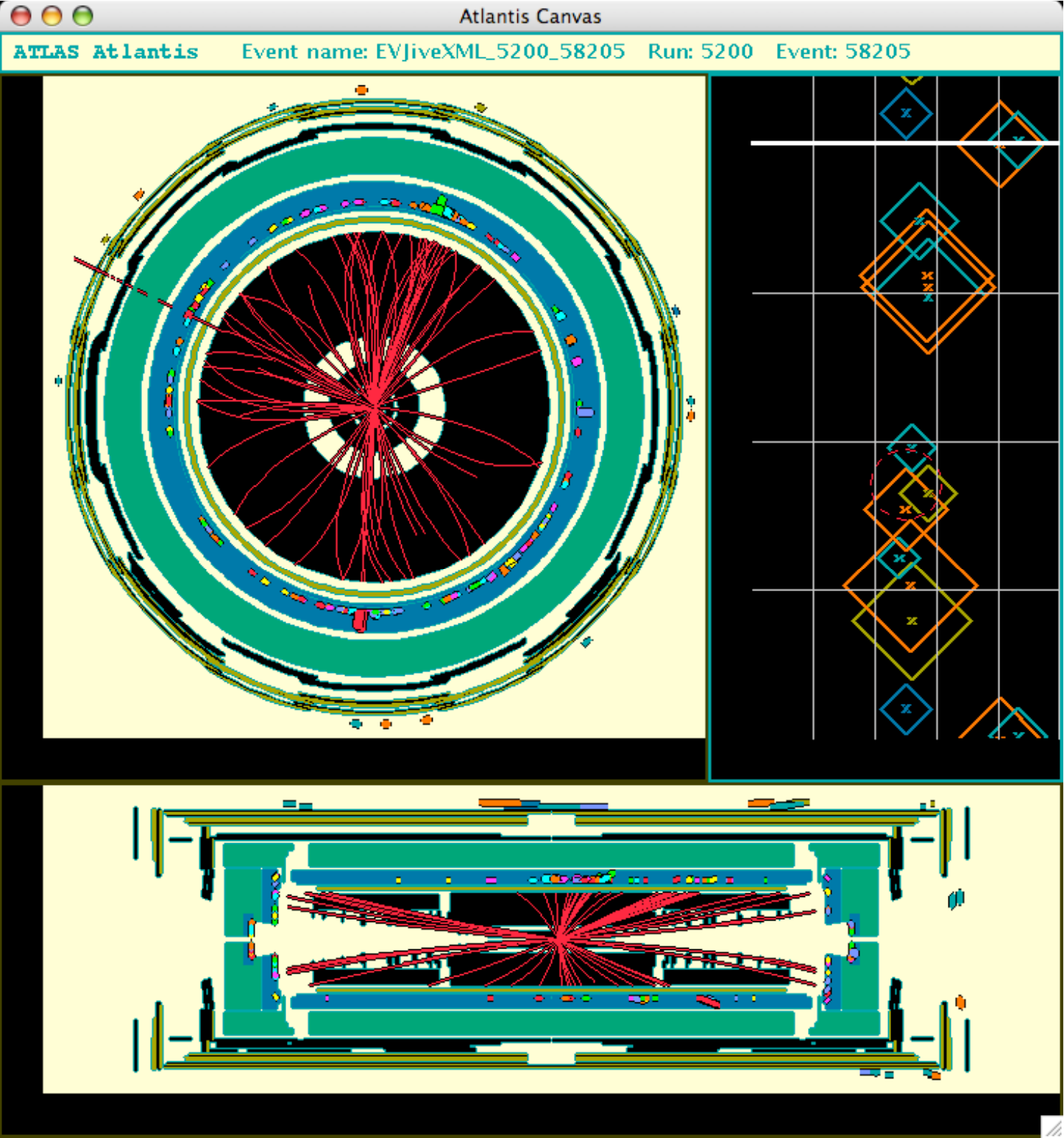} \hspace{1cm} 
	\includegraphics[width=6cm]{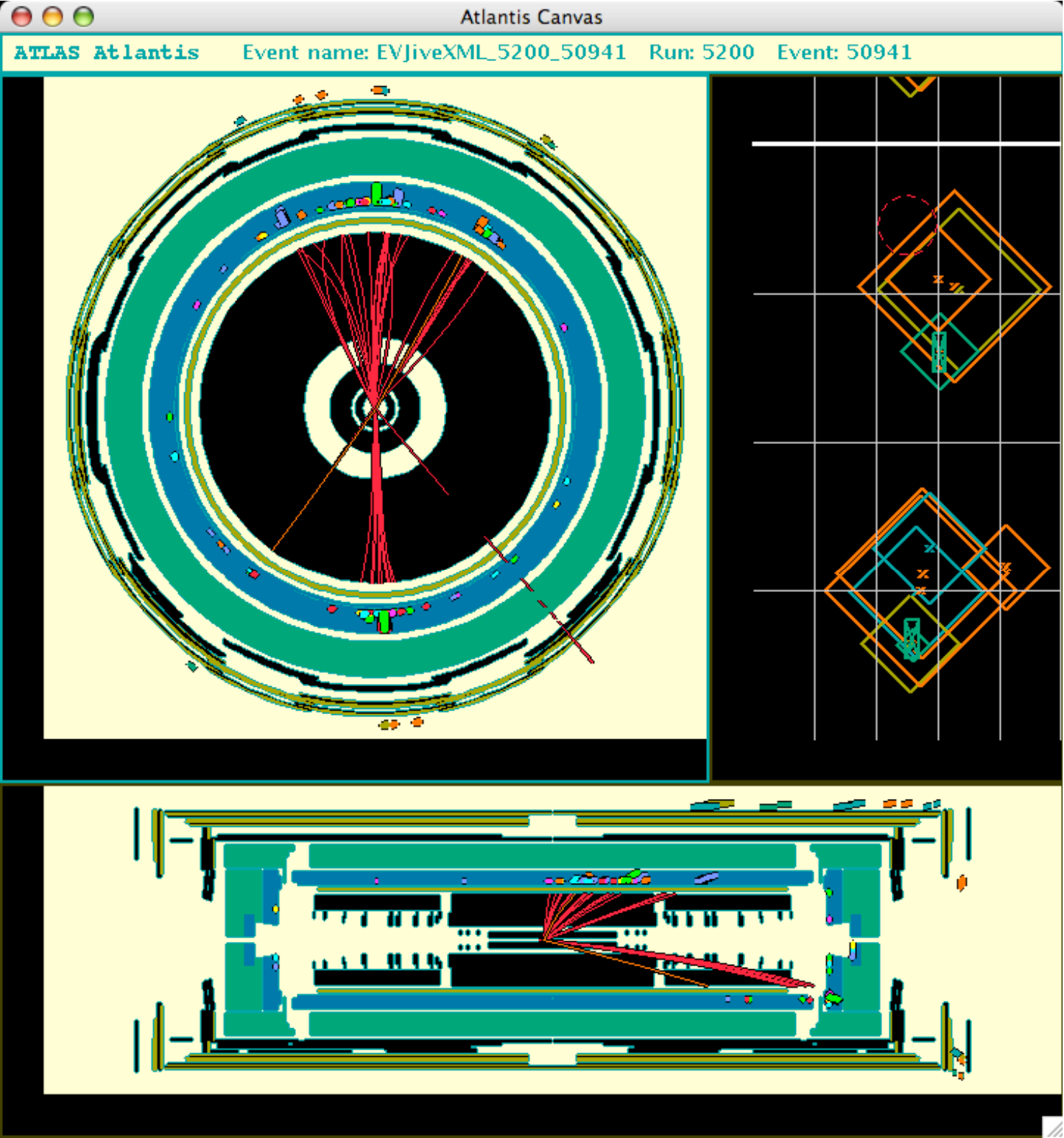} 
	\caption{Event display of \ttbar\ event where \pt\ of the top is around 150 GeV (left) and 250 GeV (right).}
	\label{Fig::atlantis}
	\end{center}
\end{figure}

At the simplest level, the measurement of the di-top system may reveal a resonance structure by fully reconstructing the \ttbar\ event. However, improvements on resolution will take more efforts. A kinematic fit can improve sensitivity in lower mass regions. On the other hand, if the resonance is located at a very high mass, the resulting top quarks will be highly boosted. Under such conditions, the top decay products start to collimate to form a single ``top-jet'' in an extreme case. 

Figure \ref{Fig::atlantis} illustrates how this occurs. While lower \pt\ top quarks spread its decay products widely in the detector making it difficult to select the correct combination of the objects, when they have a larger boost, the decay products can be more easily assigned to each top quark. With an even higher boost with \pt\ of the top above 300 GeV, however, it starts to become impossible to separate all decay products. It then becomes necessary to look into the substructure of these merged jet objects to distinguish them from high \pt\ jets originating from non-resonant QCD processes. New methods are under development to achieve this discrimination to improve the efficiency of the signal in the very high energy regime. For example, ``Y-scale'', which is used in $K_T$ jet algorithms to determine whether to merge two energy clusters, can be applied to the clusters within a jet to measure the energy scale at which the jet would split \cite{Brooijmans2008, Butterworth2002tt}. A jet containing two clusters originating from a heavy resonance would have high Y-scale while non-resonant QCD jets have much lower splitting scale. Other signatures such as displaced vertex $b$-tagging are also under investigation.

\section{Summary}
A brief overview was given to summarize the prospect for top quark physics at the LHC. The soon-to-arrive collision data will provide unmissable opportunity to the field and may lead to significant new discoveries. Both ATLAS and CMS have plans for studying the full extent of the top quark production mechanism and the top quark properties. Simulation studies have concluded that the top quark can be observed in an early stage of the experiments and the proposed new analyses methods are feasible at the LHC. It is hoped and is likely that new insights will be gained from top quarks throughout the entire lifespan of the experiments.

\bibliographystyle{atlasstylem}
\bibliography{hcps2008_conf}

\end{document}